\documentclass[epj,twocolumn]{webofc}
\usepackage[varg]{txfonts}
\usepackage{afterpage}
\usepackage{graphics}
\usepackage{fix-cm}
\usepackage{bookmark}
\usepackage{graphicx}
\usepackage{array}
\usepackage{afterpage}
\usepackage{rotating}
\usepackage{pdflscape}
\usepackage{xcolor}

\wocname{epj}
\woctitle{Seismology of the Sun and the Distant Stars 2016}
\begin{document}
\title{Stellar Parameters in an Instant with Machine Learning}
\subtitle{Application to \emph{Kepler} LEGACY Targets}
\author{\firstname{Earl P.} \lastname{Bellinger}\inst{1,2,3,4}\fnsep\thanks{\email{earl.bellinger@yale.edu}} 
   \and \firstname{George C.} \lastname{Angelou}\inst{1,3} 
   \and \firstname{Saskia} \lastname{Hekker}\inst{1,3}
   \and \firstname{Sarbani} \lastname{Basu}\inst{2}
   \and \firstname{Warrick H.} \lastname{Ball}\inst{5}
   \and \firstname{Elisabeth} \lastname{Guggenberger}\inst{1,3} }
\institute{Max-Planck-Institut f{\"u}r Sonnensystemforschung, G{\"o}ttingen, Germany
      \and Department of Astronomy, Yale University, New Haven, CT, USA 
      \and Stellar Astrophysics Centre, Department of Physics and Astronomy, Aarhus University, Denmark
      \and Institut f{\"u}r Informatik, Georg-August-Universit{\"a}t G{\"o}ttingen, Germany
      \and Institut f{\"u}r Astrophysik, Georg-August-Universit{\"a}t G{\"o}ttingen, Germany}
%
%
\abstract{
With the advent of dedicated photometric space missions, the ability to rapidly process huge catalogues of stars has become paramount. Bellinger and Angelou et al.~\cite{BA16} recently introduced a new method based on machine learning for inferring the stellar parameters of main-sequence stars exhibiting solar-like oscillations. The method makes precise predictions that are consistent with other methods, but with the advantages of being able to explore many more parameters while costing practically no time. Here we apply the method to 52 so-called ``LEGACY'' main-sequence stars observed by the \emph{Kepler} space mission. For each star, we present estimates and uncertainties of mass, age, radius, luminosity, core hydrogen abundance, surface helium abundance, surface gravity, initial helium abundance, and initial metallicity as well as estimates of their evolutionary model parameters of mixing length, overshooting coefficient, and diffusion multiplication factor. We obtain median uncertainties in stellar age, mass, and radius of 14.8\%, 3.6\%, and 1.7\%, respectively. The source code for all analyses and for all figures appearing in this manuscript can be found electronically at \url{https://github.com/earlbellinger/asteroseismology} \cite{zenodo}. 
} 
\maketitle
%
%
\section{Introduction} \label{intro}
The \emph{Kepler} seismic LEGACY sample data represents the best-quality observations of cool dwarf stars obtained during the nominal and extended 4-year mission of the \emph{Kepler} spacecraft. These stars, thought to be of a similar evolutionary stage as our Sun, serve as an excellent testbed for theories of stellar structure and evolution. In Bellinger and Angelou et al.~(hereinafter Paper 1), we introduced a method for determining the current structural parameters and evolutionary model parameters of main-sequence stars from asteroseismic observations. Here we apply that method to the LEGACY sample and present estimates of their parameters. 
%
%
\section{Data} \label{data}
\begin{figure*}
    \centering
    \includegraphics[width=\hsize]{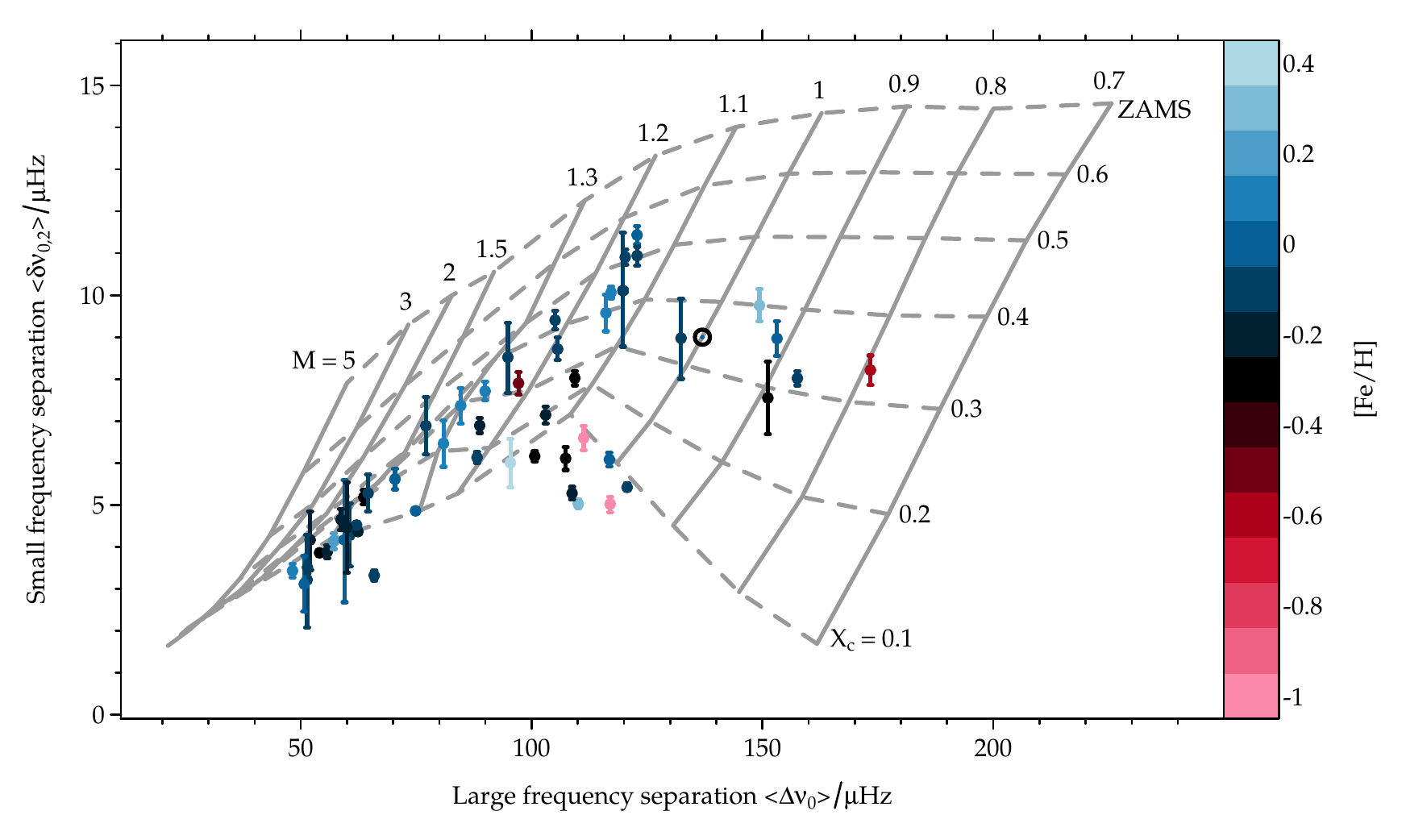}
    \caption{Small frequency separations against large frequency separations with [Fe/H] indicated by color for 52 main-sequence \emph{Kepler} LEGACY stars overplotted on top of evolutionary models varied in mass with solar-calibrated mixing length and abundances generated using MESA \cite{MESA} with frequencies calculated using GYRE \cite{GYRE}. If all stars had the solar abundances and solar mixing length, it would suffice to look up their mass and core-hydrogen abundance in this diagram. Since they do not, a more sophisticated approach is required; here we employ the method introduced in Paper 1 for this task.}
    \label{fig:jcd}
\end{figure*}

The \emph{Kepler} seismic LEGACY sample data were obtained from Lund et al.~\cite{Lund}. These data include individual frequencies, effective temperatures, frequencies of maximum oscillation power, and metallicities of 66 stars.
Although none of the LEGACY stars show mixed modes, which would be an indication that the core hydrogen burning evolutionary phase has ceased, there is no way \emph{a priori} to determine the evolutionary status of a star. Some of these stars may have already depleted their supply of core hydrogen and begun hydrogen shell burning.
As our method is currently restricted to stars on the main sequence, i.e.~stars with a fractional core hydrogen abundance $X_c \geq 10^{-3}$, we wish to only apply the method to stars that are still in this phase. 
Therefore, in order to be confident in our estimates, we adopt a very conservative inclusion criterion and do not present estimates for any stars with any part of their estimated core hydrogen distribution having $X_c \leq 10^{-2}$. 
In order to perform a selection with this criterion, we first ran our algorithm on all 66 stars. Then, for each star, we checked whether any of the $10\,000$ samples we obtained from the posterior $X_c$ distribution were smaller than that cutoff, and excluded the ones that were. 
Of the original set, 52 stars remain. The stars are visualized in an asteroseismic frequency separation diagram in Figure~\ref{fig:jcd}. 
%
%
\section{Results} \label{results} 
In Table 1 we present the means and standard deviations of current and initial parameters for 52 stars of the \emph{Kepler} LEGACY sample as inferred via the machine learning method presented in Paper 1. Figure 2 further shows the cumulative distributions of uncertainties for each of these parameters. 
\begin{figure*}
%
    \includegraphics[width=0.5\hsize]{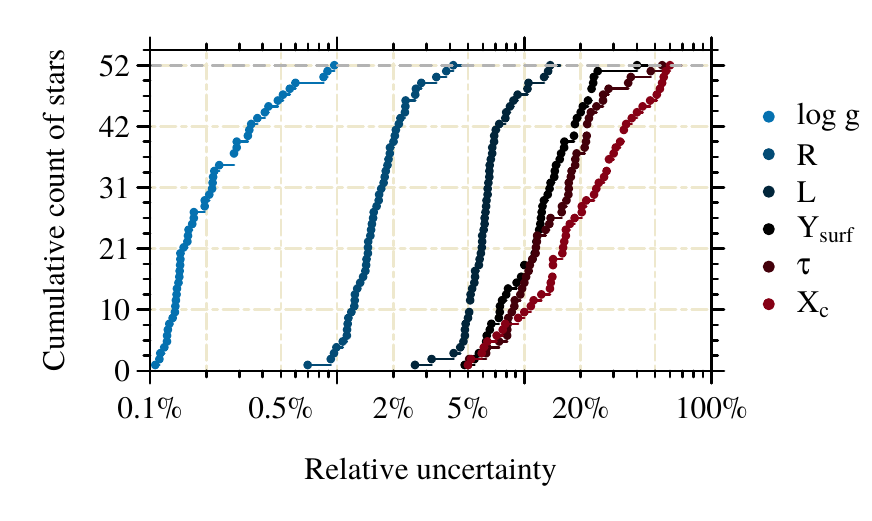}%
    \includegraphics[width=0.5\hsize]{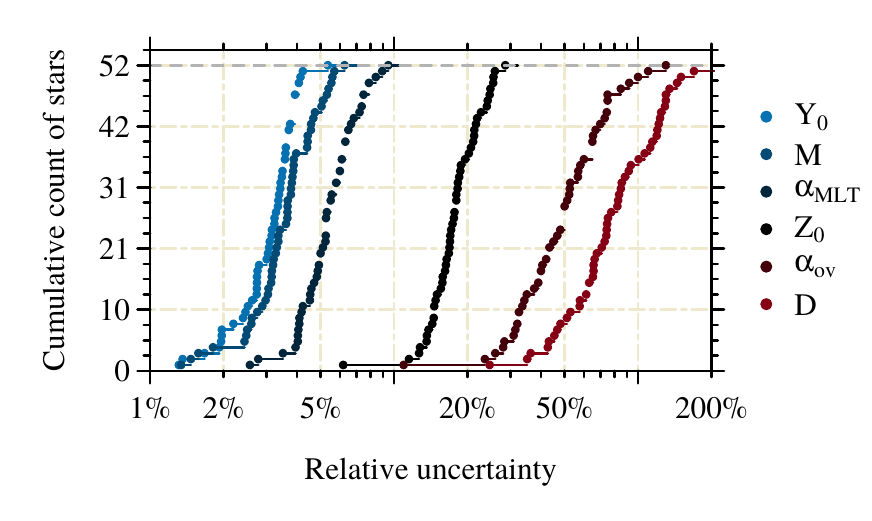}
    \caption{Cumulative distribution functions showing the relative uncertainties in estimated current parameters (left) and initial parameters (right) for 52 main-sequence \emph{Kepler} LEGACY stars.}
    \label{fig:cdf}
\end{figure*}
Nearly all of the masses are estimated to better than 5\% accuracy, with an overall average uncertainty of 3.6\%. The star with the best-constrained mass and age is KIC 8760414, an old star of 10.56 Gyr that is less massive but larger than our Sun with a mass uncertainty of 1.34\% and an age uncertainty of only 5\%. The star KIC 9139151 also has a very well-constrained age: a young star of 1.85 Gyr, its age is estimated with an uncertainty of just 210 million years. 

Surface gravity estimates are nearly an order of magnitude more precise than any other quantity, with all stars in the sample measured to better than 1\% uncertainty. In second place are radius estimates, then followed by initial helium $Y_0$. At first glance, this may seem surprising. However, our grid imposes a uniform prior in $Y_0$ spanning from $a=0.22$ to $b=0.34$. 
Thus, the largest uncertainty that would be theoretically possible is 
\begin{equation}
    \max \left( \frac{\sigma^2(Y_0)}{Y_0} \right) = \frac{|b-a|}{|a|} \cdot 100 = 54.51\%.
\end{equation}
Hence, the ``small'' relative uncertainties on $Y_0$ are actually unsurprising despite the intrinsic uncertainty in retrodicting this quantity (c.f.~Paper 1 \textsection 2.3.3). Contrast this to stellar mass, whose uncertainties are comparable, but whose maximum possible uncertainty is 128\%. 

\afterpage{
    \clearpage
    \setcounter{table}{0}
    \begin{landscape}
        \begin{table}
    \centering
    \caption{Means and standard deviations for current age $\tau$, core hydrogen abundance $X_{\mathrm{c}}$, surface gravity $\log g$, luminosity $L$, radius $R$, surface helium abundance $Y_{\mathrm{surf}}$; and initial conditions of mass $M$, initial helium $Y_0$, initial metallicity $Z_0$, mixing length parameter $\alpha_{\mathrm{MLT}}$, overshooting coefficient $\alpha_{\mathrm{ov}}$, and diffusion multiplication factor $D$ of the \emph{Kepler} Legacy data set inferred via machine learning.}
    \label{tab:results}
    \fontsize{8}{8}\selectfont
    \begin{tabular}{c|cccccc|cccccc}
        \hline\noalign{\smallskip}
        KIC & $\tau/$Gyr & $X_{\mathrm{c}}$ & $\log g$ & $L/L_\odot$ & $R/R_\odot$ & $Y_{\mathrm{surf}}$ & $M/M_\odot$ & $Y_0$ & $Z_0$ & $\alpha_{\mathrm{MLT}}$ & $\alpha_{\mathrm{ov}}$ & $D$ \\
        \noalign{\smallskip}\hline\noalign{\smallskip}
1435467         &               2.10      $\pm$     0.27            &               0.328     $\pm$     0.037           &               4.133     $\pm$     0.011           &               4.22      $\pm$     0.23            &               1.705     $\pm$     0.032           &               0.183     $\pm$     0.038           &               1.439     $\pm$     0.044           &               0.2645    $\pm$     0.0089          &               0.0262    $\pm$     0.0038          &               1.798     $\pm$     0.091           &               0.095     $\pm$     0.072           &               2.7       $\pm$     2.7             \\
3427720         &               2.8       $\pm$     1.6             &               0.44      $\pm$     0.13            &               4.376     $\pm$     0.018           &               1.56      $\pm$     0.12            &               1.123     $\pm$     0.026           &               0.223     $\pm$     0.031           &               1.090     $\pm$     0.056           &               0.280     $\pm$     0.011           &               0.0193    $\pm$     0.0038          &               1.87      $\pm$     0.14            &               0.27      $\pm$     0.17            &              15         $\pm$    11               \\
3456181         &               3.27      $\pm$     0.60            &               0.091     $\pm$     0.050           &               3.940     $\pm$     0.022           &               6.36      $\pm$     0.49            &               2.056     $\pm$     0.060           &               0.254     $\pm$     0.029           &               1.342     $\pm$     0.084           &               0.277     $\pm$     0.011           &               0.0147    $\pm$     0.0031          &               1.95      $\pm$     0.11            &               0.42      $\pm$     0.15            &               1.3       $\pm$     1.8             \\
3632418         &               3.80      $\pm$     0.71            &               0.102     $\pm$     0.058           &               4.015     $\pm$     0.024           &               4.62      $\pm$     0.32            &               1.857     $\pm$     0.049           &               0.222     $\pm$     0.037           &               1.298     $\pm$     0.071           &               0.2718    $\pm$     0.0098          &               0.0179    $\pm$     0.0039          &               1.91      $\pm$     0.13            &               0.32      $\pm$     0.15            &               3.5       $\pm$     3.6             \\
3735871         &               1.51      $\pm$     0.33            &               0.515     $\pm$     0.033           &               4.3997    $\pm$     0.0080          &               1.571     $\pm$     0.086           &               1.119     $\pm$     0.011           &               0.216     $\pm$     0.031           &               1.145     $\pm$     0.035           &               0.271     $\pm$     0.011           &               0.0208    $\pm$     0.0036          &               1.95      $\pm$     0.15            &               0.097     $\pm$     0.064           &              17         $\pm$    12               \\
5184732         &               3.86      $\pm$     0.73            &               0.246     $\pm$     0.078           &               4.266     $\pm$     0.012           &               2.01      $\pm$     0.13            &               1.374     $\pm$     0.020           &               0.254     $\pm$     0.029           &               1.271     $\pm$     0.047           &               0.274     $\pm$     0.011           &               0.0403    $\pm$     0.0068          &               1.85      $\pm$     0.16            &               0.050     $\pm$     0.045           &               2.1       $\pm$     3.5             \\
5773345         &               2.56      $\pm$     0.22            &               0.198     $\pm$     0.030           &               4.029     $\pm$     0.013           &               4.98      $\pm$     0.30            &               1.964     $\pm$     0.030           &               0.239     $\pm$     0.025           &               1.496     $\pm$     0.023           &               0.2728    $\pm$     0.0095          &               0.0329    $\pm$     0.0037          &               1.742     $\pm$     0.064           &               0.180     $\pm$     0.091           &               2.1       $\pm$     1.8             \\
6116048         &               5.96      $\pm$     0.43            &               0.080     $\pm$     0.030           &               4.2776    $\pm$     0.0051          &               1.861     $\pm$     0.087           &               1.232     $\pm$     0.013           &               0.220     $\pm$     0.020           &               1.048     $\pm$     0.028           &               0.2590    $\pm$     0.0064          &               0.0130    $\pm$     0.0018          &               1.805     $\pm$     0.074           &               0.060     $\pm$     0.018           &               2.2       $\pm$     1.4             \\
6225718         &               2.23      $\pm$     0.38            &               0.390     $\pm$     0.053           &               4.3203    $\pm$     0.0058          &               2.25      $\pm$     0.12            &               1.252     $\pm$     0.017           &               0.218     $\pm$     0.033           &               1.194     $\pm$     0.039           &               0.2730    $\pm$     0.0086          &               0.0191    $\pm$     0.0033          &               1.97      $\pm$     0.13            &               0.096     $\pm$     0.070           &               6.0       $\pm$     5.2             \\
6508366         &               3.27      $\pm$     0.58            &               0.105     $\pm$     0.053           &               3.938     $\pm$     0.034           &               6.33      $\pm$     0.60            &               2.087     $\pm$     0.081           &               0.256     $\pm$     0.026           &               1.371     $\pm$     0.079           &               0.279     $\pm$     0.010           &               0.0179    $\pm$     0.0040          &               1.99      $\pm$     0.12            &               0.46      $\pm$     0.14            &               2.7       $\pm$     3.2             \\
6603624         &               7.65      $\pm$     0.62            &               0.063     $\pm$     0.016           &               4.3355    $\pm$     0.0071          &               1.301     $\pm$     0.074           &               1.182     $\pm$     0.015           &               0.255     $\pm$     0.012           &               1.104     $\pm$     0.040           &               0.2609    $\pm$     0.0098          &               0.0327    $\pm$     0.0061          &               1.87      $\pm$     0.11            &               0.046     $\pm$     0.024           &               0.30      $\pm$     0.43            \\
6679371         &               2.54      $\pm$     0.34            &               0.109     $\pm$     0.053           &               3.935     $\pm$     0.019           &               7.38      $\pm$     0.51            &               2.159     $\pm$     0.052           &               0.271     $\pm$     0.017           &               1.461     $\pm$     0.056           &               0.280     $\pm$     0.012           &               0.0189    $\pm$     0.0029          &               1.98      $\pm$     0.11            &               0.37      $\pm$     0.14            &               0.9       $\pm$     1.2             \\
7103006         &               3.01      $\pm$     0.63            &               0.181     $\pm$     0.098           &               4.038     $\pm$     0.041           &               5.12      $\pm$     0.54            &               1.868     $\pm$     0.082           &               0.243     $\pm$     0.033           &               1.379     $\pm$     0.063           &               0.2782    $\pm$     0.0070          &               0.0228    $\pm$     0.0050          &               1.94      $\pm$     0.13            &               0.33      $\pm$     0.15            &               4.3       $\pm$     5.3             \\
7106245         &               5.64      $\pm$     0.92            &               0.143     $\pm$     0.032           &               4.3081    $\pm$     0.0091          &               1.478     $\pm$     0.065           &               1.090     $\pm$     0.014           &               0.099     $\pm$     0.040           &               0.879     $\pm$     0.021           &               0.2821    $\pm$     0.0098          &               0.0054    $\pm$     0.0012          &               1.97      $\pm$     0.13            &               0.061     $\pm$     0.039           &              12.1       $\pm$     4.3             \\
7510397         &               3.66      $\pm$     0.34            &               0.087     $\pm$     0.015           &               4.031     $\pm$     0.0040          &               4.36      $\pm$     0.28            &               1.822     $\pm$     0.038           &               0.192     $\pm$     0.022           &               1.302     $\pm$     0.058           &               0.2601    $\pm$     0.0048          &               0.0157    $\pm$     0.0026          &               1.893     $\pm$     0.094           &               0.225     $\pm$     0.068           &               3.0       $\pm$     1.3             \\
7871531         &               9.1       $\pm$     2.0             &               0.285     $\pm$     0.070           &               4.47      $\pm$     0.012           &               0.675     $\pm$     0.052           &               0.890     $\pm$     0.017           &               0.211     $\pm$     0.027           &               0.852     $\pm$     0.033           &               0.2645    $\pm$     0.0093          &               0.0137    $\pm$     0.0023          &               1.881     $\pm$     0.079           &               0.19      $\pm$     0.15            &               6.2       $\pm$     4.6             \\
7940546         &               3.47      $\pm$     0.43            &               0.068     $\pm$     0.031           &               4.0075    $\pm$     0.0088          &               4.91      $\pm$     0.32            &               1.881     $\pm$     0.036           &               0.215     $\pm$     0.027           &               1.311     $\pm$     0.058           &               0.2655    $\pm$     0.0072          &               0.0160    $\pm$     0.0027          &               1.79      $\pm$     0.10            &               0.197     $\pm$     0.080           &               1.9       $\pm$     1.4             \\
7970740         &              10.41      $\pm$     0.98            &               0.299     $\pm$     0.023           &               4.5409    $\pm$     0.0052          &               0.443     $\pm$     0.022           &               0.7770    $\pm$     0.0048          &               0.181     $\pm$     0.027           &               0.764     $\pm$     0.011           &               0.2485    $\pm$     0.0050          &               0.0088    $\pm$     0.0014          &               2.015     $\pm$     0.098           &               0.039     $\pm$     0.030           &               4.8       $\pm$     2.4             \\
8006161         &               5.1       $\pm$     1.1             &               0.412     $\pm$     0.037           &               4.4862    $\pm$     0.0088          &               0.748     $\pm$     0.046           &               0.953     $\pm$     0.016           &               0.262     $\pm$     0.016           &               1.014     $\pm$     0.031           &               0.2823    $\pm$     0.0087          &               0.0387    $\pm$     0.0078          &               2.055     $\pm$     0.096           &               0.25      $\pm$     0.13            &               3.1       $\pm$     2.5             \\
8150065         &               2.83      $\pm$     0.32            &               0.304     $\pm$     0.037           &               4.2363    $\pm$     0.0059          &               2.66      $\pm$     0.18            &               1.403     $\pm$     0.024           &               0.177     $\pm$     0.043           &               1.235     $\pm$     0.042           &               0.2680    $\pm$     0.0086          &               0.0202    $\pm$     0.0035          &               1.88      $\pm$     0.12            &               0.061     $\pm$     0.026           &               6.7       $\pm$     4.2             \\
8179536         &               1.92      $\pm$     0.90            &               0.43      $\pm$     0.12            &               4.275     $\pm$     0.017           &               2.76      $\pm$     0.17            &               1.358     $\pm$     0.025           &               0.220     $\pm$     0.042           &               1.265     $\pm$     0.069           &               0.276     $\pm$     0.011           &               0.0216    $\pm$     0.0049          &               1.88      $\pm$     0.17            &               0.15      $\pm$     0.13            &               8.9       $\pm$     9.9             \\
8228742         &               3.73      $\pm$     0.31            &               0.049     $\pm$     0.016           &               4.0231    $\pm$     0.0043          &               4.37      $\pm$     0.22            &               1.850     $\pm$     0.031           &               0.222     $\pm$     0.018           &               1.317     $\pm$     0.051           &               0.2637    $\pm$     0.0053          &               0.0192    $\pm$     0.0042          &               1.834     $\pm$     0.078           &               0.195     $\pm$     0.063           &               1.83      $\pm$     0.75            \\
8379927         &               1.43      $\pm$     0.35            &               0.523     $\pm$     0.027           &               4.3885    $\pm$     0.0091          &               1.56      $\pm$     0.13            &               1.126     $\pm$     0.016           &               0.196     $\pm$     0.035           &               1.131     $\pm$     0.050           &               0.277     $\pm$     0.015           &               0.0205    $\pm$     0.0047          &               1.90      $\pm$     0.15            &               0.133     $\pm$     0.089           &              23         $\pm$    14               \\
8394589         &               3.62      $\pm$     0.40            &               0.272     $\pm$     0.038           &               4.3255    $\pm$     0.0057          &               1.842     $\pm$     0.089           &               1.184     $\pm$     0.013           &               0.193     $\pm$     0.030           &               1.081     $\pm$     0.028           &               0.2625    $\pm$     0.0064          &               0.0130    $\pm$     0.0019          &               1.849     $\pm$     0.098           &               0.056     $\pm$     0.024           &               6.1       $\pm$     3.4             \\
8424992         &               8.98      $\pm$     0.52            &               0.096     $\pm$     0.011           &               4.3643    $\pm$     0.0050          &               1.081     $\pm$     0.049           &               1.052     $\pm$     0.018           &               0.234     $\pm$     0.018           &               0.932     $\pm$     0.031           &               0.2677    $\pm$     0.0054          &               0.0153    $\pm$     0.0031          &               1.781     $\pm$     0.066           &               0.037     $\pm$     0.013           &               1.6       $\pm$     1.1             \\
8760414         &              10.56      $\pm$     0.54            &               0.0553    $\pm$     0.0071          &               4.3194    $\pm$     0.0063          &               1.191     $\pm$     0.037           &               1.0373    $\pm$     0.0086          &               0.141     $\pm$     0.023           &               0.816     $\pm$     0.011           &               0.2599    $\pm$     0.0042          &               0.00499   $\pm$     0.00065         &               1.888     $\pm$     0.046           &               0.115     $\pm$     0.032           &               7.6       $\pm$     2.0             \\
9025370         &               4.1       $\pm$     1.5             &               0.374     $\pm$     0.075           &               4.42      $\pm$     0.015           &               0.76      $\pm$     0.11            &               1.009     $\pm$     0.024           &               0.147     $\pm$     0.037           &               0.972     $\pm$     0.047           &               0.2722    $\pm$     0.0092          &               0.0287    $\pm$     0.0072          &               1.879     $\pm$     0.087           &               0.177     $\pm$     0.099           &              30         $\pm$    13               \\
9098294         &               8.75      $\pm$     0.56            &               0.059     $\pm$     0.016           &               4.3139    $\pm$     0.0049          &               1.355     $\pm$     0.071           &               1.128     $\pm$     0.016           &               0.236     $\pm$     0.015           &               0.955     $\pm$     0.029           &               0.2630    $\pm$     0.0060          &               0.0132    $\pm$     0.0023          &               1.760     $\pm$     0.071           &               0.106     $\pm$     0.035           &               1.32      $\pm$     0.86            \\
9139151         &               1.85      $\pm$     0.21            &               0.451     $\pm$     0.032           &               4.3826    $\pm$     0.0054          &               1.93      $\pm$     0.10            &               1.167     $\pm$     0.018           &               0.260     $\pm$     0.017           &               1.199     $\pm$     0.042           &               0.2726    $\pm$     0.0099          &               0.0225    $\pm$     0.0042          &               2.08      $\pm$     0.11            &               0.084     $\pm$     0.044           &               1.6       $\pm$     2.1             \\
9139163         &               1.94      $\pm$     0.50            &               0.408     $\pm$     0.067           &               4.199     $\pm$     0.014           &               3.71      $\pm$     0.21            &               1.566     $\pm$     0.025           &               0.224     $\pm$     0.029           &               1.415     $\pm$     0.053           &               0.2754    $\pm$     0.0085          &               0.0307    $\pm$     0.0052          &               2.00      $\pm$     0.15            &               0.25      $\pm$     0.15            &               3.9       $\pm$     4.3             \\
9206432         &               1.75      $\pm$     0.39            &               0.358     $\pm$     0.074           &               4.2144    $\pm$     0.0089          &               3.69      $\pm$     0.20            &               1.496     $\pm$     0.024           &               0.274     $\pm$     0.021           &               1.339     $\pm$     0.059           &               0.291     $\pm$     0.012           &               0.0226    $\pm$     0.0042          &               1.89      $\pm$     0.18            &               0.033     $\pm$     0.035           &               0.48      $\pm$     0.92            \\
9353712         &               2.98      $\pm$     0.38            &               0.117     $\pm$     0.033           &               3.9332    $\pm$     0.0052          &               6.34      $\pm$     0.38            &               2.121     $\pm$     0.030           &               0.257     $\pm$     0.027           &               1.406     $\pm$     0.045           &               0.282     $\pm$     0.010           &               0.0190    $\pm$     0.0031          &               1.937     $\pm$     0.092           &               0.43      $\pm$     0.13            &               1.12      $\pm$     0.93            \\
9410862         &               7.17      $\pm$     0.84            &               0.072     $\pm$     0.029           &               4.3060    $\pm$     0.0059          &               1.62      $\pm$     0.094           &               1.148     $\pm$     0.019           &               0.235     $\pm$     0.016           &               0.971     $\pm$     0.034           &               0.2641    $\pm$     0.0075          &               0.0103    $\pm$     0.0018          &               1.817     $\pm$     0.095           &               0.087     $\pm$     0.047           &               1.6       $\pm$     1.3             \\
9414417         &               3.66      $\pm$     0.79            &               0.129     $\pm$     0.078           &               4.017     $\pm$     0.034           &               4.81      $\pm$     0.39            &               1.856     $\pm$     0.063           &               0.224     $\pm$     0.042           &               1.301     $\pm$     0.074           &               0.2740    $\pm$     0.0093          &               0.0179    $\pm$     0.0049          &               1.92      $\pm$     0.12            &               0.34      $\pm$     0.15            &               4.4       $\pm$     5.7             \\
9812850         &               2.46      $\pm$     0.43            &               0.263     $\pm$     0.065           &               4.095     $\pm$     0.021           &               4.53      $\pm$     0.31            &               1.765     $\pm$     0.045           &               0.185     $\pm$     0.041           &               1.412     $\pm$     0.054           &               0.266     $\pm$     0.011           &               0.0235    $\pm$     0.0045          &               1.83      $\pm$     0.11            &               0.14      $\pm$     0.10            &               4.5       $\pm$     5.7             \\
9955598         &               7.2       $\pm$     1.2             &               0.363     $\pm$     0.039           &               4.4916    $\pm$     0.0077          &               0.662     $\pm$     0.041           &               0.902     $\pm$     0.013           &               0.246     $\pm$     0.020           &               0.920     $\pm$     0.029           &               0.270     $\pm$     0.010           &               0.0223    $\pm$     0.0038          &               1.940     $\pm$     0.092           &               0.22      $\pm$     0.14            &               3.0       $\pm$     2.2             \\
9965715         &               3.49      $\pm$     0.59            &               0.281     $\pm$     0.049           &               4.269     $\pm$     0.012           &               1.96      $\pm$     0.18            &               1.267     $\pm$     0.028           &               0.140     $\pm$     0.033           &               1.068     $\pm$     0.048           &               0.2814    $\pm$     0.0086          &               0.0154    $\pm$     0.0036          &               1.709     $\pm$     0.073           &               0.080     $\pm$     0.036           &              15.8       $\pm$     7.5             \\
10068307        &               3.91      $\pm$     0.29            &               0.071     $\pm$     0.014           &               3.9423    $\pm$     0.0054          &               5.24      $\pm$     0.29            &               2.012     $\pm$     0.020           &               0.225     $\pm$     0.027           &               1.287     $\pm$     0.028           &               0.2720    $\pm$     0.0035          &               0.0145    $\pm$     0.0018          &               1.900     $\pm$     0.066           &               0.431     $\pm$     0.048           &               2.5       $\pm$     1.8             \\
10079226        &               2.21      $\pm$     0.62            &               0.447     $\pm$     0.065           &               4.3701    $\pm$     0.0097          &               1.57      $\pm$     0.10            &               1.174     $\pm$     0.022           &               0.217     $\pm$     0.033           &               1.177     $\pm$     0.046           &               0.271     $\pm$     0.011           &               0.0284    $\pm$     0.0050          &               1.90      $\pm$     0.15            &               0.13      $\pm$     0.11            &              14         $\pm$    12               \\
10162436        &               3.43      $\pm$     0.29            &               0.085     $\pm$     0.030           &               3.9788    $\pm$     0.0072          &               5.07      $\pm$     0.29            &               1.970     $\pm$     0.028           &               0.202     $\pm$     0.026           &               1.345     $\pm$     0.042           &               0.2696    $\pm$     0.0047          &               0.0182    $\pm$     0.0025          &               1.896     $\pm$     0.076           &               0.280     $\pm$     0.096           &               3.7       $\pm$     1.8             \\
10454113        &               1.36      $\pm$     0.32            &               0.490     $\pm$     0.030           &               4.3257    $\pm$     0.0064          &               2.11      $\pm$     0.12            &               1.263     $\pm$     0.015           &               0.163     $\pm$     0.035           &               1.229     $\pm$     0.029           &               0.2786    $\pm$     0.0094          &               0.0251    $\pm$     0.0038          &               2.00      $\pm$     0.12            &               0.098     $\pm$     0.066           &              25         $\pm$     11              \\
10644253        &               1.06      $\pm$     0.28            &               0.562     $\pm$     0.034           &               4.4029    $\pm$     0.0070          &               1.525     $\pm$     0.089           &               1.123     $\pm$     0.016           &               0.234     $\pm$     0.026           &               1.163     $\pm$     0.044           &               0.277     $\pm$     0.011           &               0.0239    $\pm$     0.0038          &               1.85      $\pm$     0.13            &               0.093     $\pm$     0.055           &              16         $\pm$     12              \\
10730618        &               4.16      $\pm$     0.80            &               0.171     $\pm$     0.024           &               4.0547    $\pm$     0.0087          &               4.01      $\pm$     0.52            &               1.760     $\pm$     0.037           &               0.206     $\pm$     0.050           &               1.275     $\pm$     0.071           &               0.2764    $\pm$     0.0077          &               0.0199    $\pm$     0.0053          &               1.994     $\pm$     0.099           &               0.42      $\pm$     0.11            &               5.8       $\pm$     4.7             \\
10963065        &               4.35      $\pm$     0.46            &               0.158     $\pm$     0.044           &               4.2929    $\pm$     0.0062          &               1.998     $\pm$     0.099           &               1.234     $\pm$     0.014           &               0.224     $\pm$     0.026           &               1.089     $\pm$     0.030           &               0.2633    $\pm$     0.0076          &               0.0136    $\pm$     0.0024          &               1.81      $\pm$     0.11            &               0.052     $\pm$     0.023           &               2.4       $\pm$     1.8             \\
11081729        &               1.69      $\pm$     0.27            &               0.428     $\pm$     0.033           &               4.2537    $\pm$     0.0051          &               3.46      $\pm$     0.18            &               1.446     $\pm$     0.019           &               0.247     $\pm$     0.020           &               1.368     $\pm$     0.044           &               0.2739    $\pm$     0.0081          &               0.0249    $\pm$     0.0044          &               2.11      $\pm$     0.14            &               0.118     $\pm$     0.059           &               1.8       $\pm$     2.5             \\
11253226        &               1.75      $\pm$     0.60            &               0.456     $\pm$     0.066           &               4.19      $\pm$     0.020           &               4.34      $\pm$     0.28            &               1.567     $\pm$     0.028           &               0.224     $\pm$     0.040           &               1.388     $\pm$     0.069           &               0.2735    $\pm$     0.0092          &               0.0214    $\pm$     0.0054          &               1.90      $\pm$     0.11            &               0.35      $\pm$     0.17            &               3.7       $\pm$     4.8             \\
11772920        &               9.08      $\pm$     0.92            &               0.305     $\pm$     0.014           &               4.4958    $\pm$     0.0064          &               0.497     $\pm$     0.067           &               0.8635    $\pm$     0.0095          &               0.202     $\pm$     0.048           &               0.851     $\pm$     0.026           &               0.265     $\pm$     0.011           &               0.0208    $\pm$     0.0058          &               1.92      $\pm$     0.10            &               0.124     $\pm$     0.079           &               6.2       $\pm$     5.6             \\
12009504        &               3.64      $\pm$     0.32            &               0.181     $\pm$     0.029           &               4.2147    $\pm$     0.0053          &               2.72      $\pm$     0.13            &               1.424     $\pm$     0.017           &               0.213     $\pm$     0.028           &               1.212     $\pm$     0.037           &               0.2649    $\pm$     0.0080          &               0.0177    $\pm$     0.0024          &               1.90      $\pm$     0.11            &               0.047     $\pm$     0.022           &               2.2       $\pm$     1.5             \\
12069127        &               2.71      $\pm$     0.30            &               0.106     $\pm$     0.048           &               3.9052    $\pm$     0.0071          &               7.03      $\pm$     0.41            &               2.232     $\pm$     0.026           &               0.278     $\pm$     0.018           &               1.460     $\pm$     0.037           &               0.292     $\pm$     0.010           &               0.0234    $\pm$     0.0035          &               1.867     $\pm$     0.090           &               0.41      $\pm$     0.15            &               0.57      $\pm$     0.57            \\
12069449        &               6.89      $\pm$     0.36            &               0.154     $\pm$     0.027           &               4.3547    $\pm$     0.0064          &               1.234     $\pm$     0.038           &               1.115     $\pm$     0.012           &               0.244     $\pm$     0.013           &               1.024     $\pm$     0.015           &               0.2675    $\pm$     0.0049          &               0.0211    $\pm$     0.0014          &               1.791     $\pm$     0.048           &               0.103     $\pm$     0.033           &               1.34      $\pm$     0.73            \\
12258514        &               4.25      $\pm$     0.34            &               0.0677    $\pm$     0.0097          &               4.1225    $\pm$     0.0050          &               3.06      $\pm$     0.15            &               1.618     $\pm$     0.021           &               0.217     $\pm$     0.027           &               1.264     $\pm$     0.043           &               0.2646    $\pm$     0.0073          &               0.0221    $\pm$     0.0032          &               1.85      $\pm$     0.10            &               0.100     $\pm$     0.024           &               2.7       $\pm$     1.9             \\
12317678        &               2.73      $\pm$     0.28            &               0.094     $\pm$     0.035           &               4.0494    $\pm$     0.0052          &               5.39      $\pm$     0.34            &               1.781     $\pm$     0.029           &               0.262     $\pm$     0.019           &               1.297     $\pm$     0.047           &               0.2767    $\pm$     0.0087          &               0.0114    $\pm$     0.0015          &               1.85      $\pm$     0.11            &               0.155     $\pm$     0.091           &               0.20      $\pm$     0.23            \\
Sun             &               4.62      $\pm$     0.17            &               0.355     $\pm$     0.012           &               4.4393    $\pm$     0.0042          &               1.015     $\pm$     0.043           &               1.0004    $\pm$     0.0064          &               0.2458    $\pm$     0.0077          &               1.0022    $\pm$     0.0095          &               0.2665    $\pm$     0.0035          &               0.0182    $\pm$     0.0011          &               1.819     $\pm$     0.036           &               0.072     $\pm$     0.023           &               2.14      $\pm$     0.72            \\
    \noalign{\smallskip}\hline
    \multicolumn{7}{l}{\textbf{Note.} The values obtained from degraded solar data (c.f.\ Paper 1 \textsection 3.2) predicted on these quantities are shown for reference.}
    \end{tabular}
\end{table}
    \end{landscape}
    \clearpage
}

In Table 2, we show these maximum theoretically possible uncertainties for all twelve quantities that we estimate. We compare them with the actual average uncertainties obtained across the 52 stars analyzed here. We also calculate a truncated average explained variance score
\begin{equation}
    V_{\text{e, mean}}^{\text{trunc}} = 1 - \frac{ 
            \sqrt{\sigma^2_{\text{mean}}} 
        } { 
            \sqrt{\min \left( \sigma^2_{\max}, 200 \right)}
    }
\end{equation} (c.f.~Paper 1 equation 8), which gives an indication of how well the predictions compare with a random guess, with a score of zero being no better and a score of one being much better. We truncate at 200 because quantities that can take on a value of zero otherwise vacuously give $V_{\text{e, mean}}=1$. 

\setcounter{table}{1}
\begin{table}
    \centering
    \caption{Maximum theoretically possible uncertainties, average uncertainties attained, and truncated average explained variance of the twelve estimated quantities for the 52 \emph{Kepler} LEGACY stars.}
    \label{tab:uncertainties}
    \begin{tabular}{c|ccc}
        \hline\noalign{\smallskip}
Quantity              & $  \sigma^2_{\max}$ & $\sigma^2_{\text{mean}}$ & $V_{\text{e, mean}}^{\text{trunc}}$ \tabularnewline \hline
$\tau$                & $           \infty$ & $16.76\%$ & 0.710 \tabularnewline
$X_c$                 & $32\,612\,260   \%$ & $23.98\%$ & 0.653 \tabularnewline
$\log g$              & $          56.42\%$ & $ 0.26\%$ & 0.932 \tabularnewline
$L$                   & $     71\,054.89\%$ & $ 6.59\%$ & 0.818 \tabularnewline
$R$                   & $         919.52\%$ & $ 1.74\%$ & 0.906 \tabularnewline
$Y_{\text{surf}}$     & $           \infty$ & $13.58\%$ & 0.739 \tabularnewline \hline
$M$                   & $         128.48\%$ & $ 3.67\%$ & 0.830 \tabularnewline
$Y_0$                 & $          54.51\%$ & $ 3.14\%$ & 0.759 \tabularnewline
$Z_0$                 & $    999\,060   \%$ & $18.19\%$ & 0.698 \tabularnewline
$\alpha_{\text{MLT}}$ & $          66.65\%$ & $ 5.60\%$ & 0.710 \tabularnewline
$\alpha_{\text{ov}}$  & $           \infty$ & $53.29\%$ & 0.483 \tabularnewline
$D$                   & $           \infty$ & $86.86\%$ & 0.340 \tabularnewline \noalign{\smallskip}\hline
    \multicolumn{4}{p{0.9\linewidth}}{
    \raggedright\textbf{Note.} Quantities that can take on a value of zero, such as a ZAMS star with $\tau=0$, have $\sigma^2_{\max}=\infty$. 
    }
    \end{tabular}
\end{table}

Based on these scores, the most well-constrained parameters are $\log g$, $R$, and $M$. The parameters that are hardest to constrain are the $\alpha_{\text{ov}}$ and $D$. All of the stars have $\sigma^2(D)/D > 0.2$ and nearly a third of them are more than 100\% uncertain. This highlights several aspects. First, since $D$ can take on a value of zero, an infinite relative uncertainty is possible. Second, $D$ is highly degenerate with the parameters controlling the initial chemical composition. These uncertainties may merely represent that degeneracy. Third, there are mixing processes that are not correctly accounted for in one-dimensional stellar modelling. Extreme values of $D$ may therefore be compensating for those processes. Finally, there may be seismic diagnostics, e.g.~glitch analysis, that would be able to better constrain diffusion, but are absent from the present analysis.

\section{Conclusions}
\label{conclusions}

In this paper, we applied machine learning techniques to estimate structural and evolutionary parameters of main-sequence stars. We achieved extremely precise estimates of stellar mass and radius using asteroseismologic diagnostics that are competitive with orbital modelling and even direct interferometric measurements. Hence, these estimates represent one of the largest and most precise collections of main-sequence stellar parameters. 

There are other modelling efforts of this LEGACY sample. Silva Aguirra et al.~\cite{SA} applied seven different techniques based on iterative optimization to estimate the parameters of these stars. 
Although we have shown in Paper 1 that the results are in good agreement, the philosophy of our approach is fundamentally different from those seven. 
Those approaches are based on various strategies for searching through grids of models in order to optimize a goodness-of-fit criterion. 
Our approach, which is based on classification and regression trees (CART), works without searching and essentially without the tuning of hyper-parameters. 
Moreover, our approach enables estimation of many more stellar parameters, such as the initial helium abundance, mixing length parameter, overshooting coefficient, and diffusion multiplication factor, which would be too computationally expensive to vary with search-based methods, while still only taking seconds per star. 

We have omitted several stars due to their proximity to the end of the main sequence. We are currently working on extending this method to more evolved stellar types, and we are soon to release a follow-up paper analyzing these omitted stars as well as more evolved ones. 

\textbf{Acknowledgements.} The research leading to the presented results has received funding from the European Research Council under the European Community's Seventh Framework Programme (FP7/2007-2013) / ERC grant agreement no 338251 (StellarAges). This research was undertaken in the context of the International Max Planck Research School for Solar System Research. S.B.\ acknowledges partial support from NSF grant AST-1514676 and NASA grant NNX13AE70G. W.H.B.\ acknowledges research funding by Deutsche Forschungsgemeinschaft (DFG) under grant SFB 963/1 ``Astrophysical flow instabilities and turbulence'' (Project A18).

\textbf{Software.} Analysis was performed with R 3.2.3 \citep{R} and the R libraries magicaxis 1.9.4 \citep{magicaxis}, parallelMap 1.3 \citep{parallelMap}, and matrixStats 0.50.1 \citep{matrixStats}.

\end{document}